\documentclass[american,aps,pra,reprint,superscriptaddress,twocolumn]{revtex4-1}
\usepackage[unicode=true,pdfusetitle, bookmarks=true,bookmarksnumbered=false,bookmarksopen=false, breaklinks=false,pdfborder={0 0 0},backref=false,colorlinks=false] {hyperref}
\hypersetup{ colorlinks,linkcolor=myurlcolor,citecolor=myurlcolor,urlcolor=myurlcolor}
\usepackage{braket,colortbl,cleveref,amsthm,amsmath,amssymb,txfonts}
\definecolor{myurlcolor}{rgb}{0,0,0.7}
\usepackage{graphics,graphicx}
\usepackage{color}

\theoremstyle{plain}

\def\bea{\begin{eqnarray}}
\def\eea{\end{eqnarray}}
\def\ba{\begin{array}}
\def\ea{\end{array}}

\def\ket{\rangle}
\def\bra{\langle}
\def\beq{\begin{equation}}
\def\eeq{\end{equation}}

\begin{document}

\title{ Universal quantum uncertainty relations between non-ergodicity and loss of information}

\author{Natasha Awasthi}
\affiliation{ College of Basic Sciences and Humanities, G.B. Pant University Of Agriculture and Technology, Pantnagar, Uttarakhand - 263153, India \\ }
\affiliation{Harish-Chandra Research Institute, HBNI, Chhatnag Road, Jhunsi, Allahabad - 211019, India}

\author{Samyadeb Bhattacharya} 
\affiliation{Harish-Chandra Research Institute, HBNI, Chhatnag Road, Jhunsi, Allahabad - 211019, India} 
\affiliation{S. N. Bose National Centre for Basic Sciences, Block JD, Sector III, Salt Lake, Kolkata 700 098, India}

\author{Aditi Sen(De)} 

\author{Ujjwal Sen}

\affiliation{Harish-Chandra Research Institute, HBNI, Chhatnag Road, Jhunsi, Allahabad - 211019, India}


\begin{abstract}
\noindent We establish uncertainty relations between information loss in general open quantum systems and the amount of non-ergodicity of the corresponding dynamics. The relations hold for arbitrary quantum systems interacting with an arbitrary quantum environment. The elements of the uncertainty relations are quantified via distance measures on the space of quantum density matrices. 
The relations hold for arbitrary distance measures satisfying a set of intuitively satisfactory axioms. The relations show that as the non-ergodicity of the dynamics increases, the lower bound on information loss decreases, which validates the belief that non-ergodicity plays an important role in preserving information of quantum states undergoing lossy evolution. We  also consider a model of a central qubit  interacting with a fermionic thermal bath and derive its reduced dynamics, to subsequently investigate the information loss and non-ergodicity in such dynamics. We comment on the ``minimal'' situations that saturate the uncertainty relations.

\end{abstract}

\maketitle

\section{Introduction}

In practical situations, it is arguably impossible to completely isolate a quantum system from its surroundings and it is subjected to information loss due to dissipation and decoherence. In modelling open quantum systems, the simpler approach is to consider the environment to be memoryless, i.e. Markovian \citep{Lindblad,gorini,breuer,alicki,wilde}. The system-environment relation is however more often than not non-Markovian,
and there are possibilities of information backflow into the system, which can be considered as a resource in information theoretic tasks \citep{NM1,NM2,NM3}. The systems showing such properties are usually associated with various structured environments without the consideration of weak system-environment coupling and the Born- Markov approximation \citep{central2,spinbath1,spinbath2,central3,central5,central6,NM4,samya2,samya1}. 
In a Markovian evolution, this information flow is one-way and quickly leads to an unwanted total loss of coherence and other quantum characteristics. Using structured environments, it may be possible to reduce information loss of the associated quantum system. 

On the other hand, an important statistical mechanical attribute of a  system interacting with an environment, with the later being in a thermal
state, is the ergodicity of the system.  A physical process is considered to be ergodic, if the statistical properties of the process can be realized from a long-time averaged realization. In the study of the realization of a thermal relaxation process, ergodicity plays a very important role \citep{breuer,ergodicity1,ergodicity2}. 
It  also has important applications in quantum control \citep{control1,control2,control3,control4}, quantum communication \citep{com1}, and beyond \citep{other1,other2}. Here we intend to capture the notion of ``non-ergodicity" from the perspective of quantum channels, i.e. considering only the reduced dynamics of a quantum system interacting with an environment. In the framework of open quantum systems, a rigorous study on ergodic quantum channels can be found in \citep{ergodicity3}. Ergodic quantum channels are channels having a unique fixed point in the space of density matrices \citep{ergodicity4}. Non-ergodicity of a dynamical process can then be quantified as the amount of deviation from a ergodic process in open system dynamics. 

In this work, we find a connection between information loss of a general open quantum system and non-ergodicity therein. We propose a  measure of information loss in a quantum system, based on distinguishability
 of quantum states, which in turn is based on distance measures on the space of density operators \citep{blp1,alonso,breuerN,breuerN1,rivas1,sabrina1}. We quantify the non-ergodicity of the dynamics based on the distance between the time-averaged state after sufficiently long processing time and the corresponding thermal equilibrium state. Within this paradigm, we derive an uncertainty
 relation between informaton loss and the amount of non-ergodicity for an arbitrary quantum system interacting according to an arbitrary quantum Hamiltonian with an arbitrary
 environment. The derivation is not for a particular distance measure, but for all such which satisfies a set of intuitively satisfactory axioms. In the illustrations, we mainly focus on the 
 trace distance, and to a certain extent, also on the relative entropy. We find that our  relations  are compatible with Markovian ergodic dynamics, where the system loses all the information.  

Finally, we have considered a particular structured environment model, where a central qubit interacts with a collection of mutually non-interacting spins in thermal states at an arbitrary temperature. A spin-bath model of this type, which has been considered previously in the literature \citep{samya1,samya2,spinbath1,spinbath2,spinRev}, shows a  highly non-Markovian nature. Here we have derived the reduced dynamics of  a particular spin-bath model without the weak coupling and Born-Markov approximations. Subsequently, we investigate the information loss and non-ergodicity, and find
the status of the uncertainty for this system. 

The organization of the paper is as follows. In Section II, we present the definitions for loss of information and non-ergodicity. We derive the uncertainty relations between information loss and non-ergodicity in Section III. In Section IV, we consider the central spin model, derive the reduced dynamics of the central qubit, and analyze the corresponding information loss and non-ergodicity. We conclude in Section V.

\section{Definitions: Measures for Loss Of Information  And  Non-ergodicity}

Before proceeding to the main results, let us  define the two primary quantities under present investigation, i.e. loss of information and a measure of non-ergodicity, based on distance measures. 

\subsection{Loss of information}
We quantify the loss of information in quantum systems due to environmental interaction, in terms of distinguishability measures  for quantum states. The loss of information, denoted by $I_{\Delta}(t),$ at any instant of  time, can be quantified by the maximal difference between the initial distinguishability between a pair of states, $\rho_1(0),~ \rho_2(0)$, and that for the corresponding time evolved states $\rho_1(t)=\Phi(\rho_1(0)),~ \rho_2(t)=\Phi(\rho_2(0))$ at time t, where $\Phi$ denotes the open quantum evolution of the initial states.
Mathematically, it is given by
\beq\label{sec1a}
I_{\Delta}(t)= \max_{\rho_1(0),~\rho_2(0)} \left(D(\rho_1(0),\rho_2(0))-D(\rho_1(t),\rho_2(t)) \right),
\eeq
where the distance measure $D(\rho,\sigma)$ must satisfy the following conditions: 
\begin{enumerate}
\item[P1.] $D(\rho,\sigma)\geq 0~~\forall~~\mbox{density matrices}~~\rho, \sigma$.
\item[P2.] $D(\rho,\rho)=0~~\forall~~\rho~$  and  $~D(\rho,\sigma)=0~~\Longleftrightarrow~~\rho=\sigma,~~\forall~~\rho,\sigma$.
\item[P3.] $D(\Phi(\rho),\Phi(\sigma))\leq D(\rho,\sigma)~~\forall~~\rho,\sigma~~ $and $\forall~~$ completely positive trace preserving maps $\Phi(\cdot)$, on the space of density operators, $\mathcal{B}(\mathcal{H})$, on the Hilbert space $\mathcal{H}$.
\end{enumerate}
The class of distance measures satisfying these conditions, include
 trace distance, Bures distance, Hellinger distance \citep{dist1,dist2,dist4}. Though the von Neumann relative entropy and Jensen-Shannon divergence also satisfy the aforementioned conditions, they are not generally considered as geometric distances, since they certain other metric properties.
But also note here that the square root of Jensen-Shannon divergence does satisfy metric properties \citep{dist3,metric1,metric2} and can be considered as a valid distance measure. It is also important to mention that all the aforementioned valid distance measures are bounded. 

\textbf{Loss of information for time-averaged states:} To draw the connection with non-ergodicity, discussed below, we now define the long time-averaged state as 
\beq\label{sec1b}
\bar{\rho}= \lim_{\tau\rightarrow\infty} \frac{1}{\tau}\int_0^{\tau}\rho(t)dt .
\eeq
The information loss for the time-averaged state, which we call `` average loss of information'', can then be defined as 
\beq\label{sec1c}
\bar{I}_{\Delta}= \max_{\rho_1(0),~\rho_2(0)} \left(D(\rho_1(0),\rho_2(0))-D(\bar{\rho}_1,\bar{\rho}_2) \right),
\eeq
which is lower and upper bounded by 0  and 1 respectively. From Eq. \eqref{sec1c} we can infer that, when the open system dynamics has a unique steady state or fixed point, independent of initial state, the entire information $D(\rho_1(0),\rho_2(0))$ is lost for arbitrary inputs $\rho_1(0),\rho_2(0)$. Later in the paper, we draw a connection between average loss of information and non-ergodicity of the underlying dynamics,
with the later being defined in the succeeding subsection.
\subsection{Non-ergodicity}
Ergodicity plays an important role in  statistical mechanics, to describe the realization of relaxation of the system to the thermal equilibrium.
The ergodic hypothesis states that if a system evolves over a long period of time, 
 the long time-averaged state of the system is equal to its thermal state corresponding to the temperature of the environment with which the system is interacting. Ergodicity can also be defined in terms of observables. For any observable $f$, if its long time average, $\bra f\ket_T$ is equal to its ensemble average, $\bra f\ket_{en}$, the dynamics is considered to be ergodic for the observables. Here the time and ensemble averages of the observable are respectively defined as 
 \[
 \bra\bar{f}\ket =\lim_{\tau\to\infty}\frac{1}{\tau}\int_0^{\tau} \mbox{Tr}[f\rho(t)] = \mbox{Tr}[f\bar{\rho}]~~;~~ \bra f\ket_{en} = \mbox{Tr}[f\rho_{th}].
 \]
 Ergodicity further assumes the equality of $\bra\bar{f}\ket$ and $\bra{f}\ket_{en}$, independent of the initial state of the evolution.  
 Therefore, non-ergodicity of the dynamics for the observable can be quantified by the difference between the time average and ensemble average, i.e. by $|\bra\bar{ f}\ket-\bra f\ket_{en}|= |\mbox{Tr}[f (\bar{\rho}-\rho_{th})]|$. 
 Based on these understandings of ergodicity of a dynamics, we define a measure of non-ergodicity as the distance between the long time-averaged state ($\bar{\rho}$) of the system and its corresponding thermal state ($\rho_{th}$), and so is given by
\beq\label{sec1d}
\mathcal{N}_{\epsilon}(\bar{\rho}) =  D(\bar{\rho}~,~\rho_{th}).
\eeq
Here we  impose  two further conditions on the allowed distance measures:
\begin{enumerate}
\item[P4.] The measure must be symmetric, i.e $D(\rho,\sigma)=D(\sigma, \rho)$,~~ $\forall \rho,\sigma$.
\item[P5.] The measure must satisfy the triangle inequality, given by $D(\rho,\sigma) \leq D(\rho,\kappa)+D(\kappa,\sigma),~~\forall~~\mbox{density matrices}~~ \rho, \sigma,\kappa$.
\end{enumerate}
The conditions P1-P5 are satisfied by the geometric distance measures like trace distance, Bures distance, and Hellinger  distance. Note the von Neumann relative entropy neither satisfies the symmetry property nor
the triangle inequality and hence we cannot use it directly for our investigation. However, we will later show the possibility of overcoming such ``shortcomings'' of the relative 
entropy distance. Interestingly, it has been shown  \citep{metric1} that Jenson-Shannon divergence satisfies the symmetry property and for its square root, the triangle inequality holds. Therefore,
the square root of Jensen-Shannon divergence can also be taken as a proper distance measure for our investigation. Note that the measure of non-ergodicity, given in \eqref{sec1d}, depends on the initial state.
Hence, to  obtain a measure of  non-ergodicity which is state-independent, we introduce
\beq\label{sec1dg}
\mathcal{N}_{\epsilon}^M =\max_{\rho(0)}\mathcal{N}_{\epsilon}(\bar{\rho}),
\eeq
where maximization is performed over all initial states ($\rho(0)$).
\section{Connecting information loss with non-ergodicity}
With the definitions given in the preceding section, we now establish a connection between loss of information and non-ergodicity. For the distance measures, which satisfy P1-P5, we obtain
\[ D(\bar{\rho}_1,\bar{\rho}_2) \leq \mathcal{N}_{\epsilon}(\bar{\rho}_1)+\mathcal{N}_{\epsilon}(\bar{\rho}_2). \] 
Using Eq. (\ref{sec1c}), we therefore have the inequality
\beq\label{sec1e}
\bar{I}_{\Delta} \geq \max_{\rho_1(0),~\rho_2(0)} \left(D(\rho_1(0),\rho_2(0))-\left(\mathcal{N}_{\epsilon}(\bar{\rho}_1)+\mathcal{N}_{\epsilon}(\bar{\rho}_2)\right)\right).
\eeq
It draws a direct connection between non-ergodicity and loss of information in open system dynamics. 
Using the state-independent measure of non-ergodicity (Eq. \eqref{sec1dg}), we can arrive at an uncertainty relation between information loss and
a measure of non-ergodicity, given by 
\beq\label{sec1eg}
\bar{I}_{\Delta} + 2\mathcal{N}_{\epsilon}^M \geq \max_{\rho_1(0),~\rho_2(0)} \left(D(\rho_1(0),\rho_2(0))\right).
\eeq
The above relation is valid for any distance measure which satisfies the conditions P1-P5, and for any quantum system, interacting with an arbitrary
environment.

In this paper, we will mainly work  on the uncertainty relation based on the distance measure given by $D^T(\rho,\sigma)=\frac{1}{2}\mbox{Tr}|\rho-\sigma|$ for pairs of states $\rho$ and $\sigma$.
The importance of quantum relative entropy \citep{vedral,Ndutta1,NDutta2} as a ``distance-type'' measure, notwithstanding its inability in satisfying symmetry and other relations, from the perspective of quantum thermodynamics is unquestionable, and hence obtaining uncertainity relation in terms of 
quantum relative entropy can be interesting. Towards this aim, we use 
a relation between relative entropy and trace distance \citep{relEnt}, given by
\beq\label{new1}
S(\rho||\sigma)\equiv\mbox{Tr}[\rho(\log\rho-\log\sigma)]\geq 2(D^T(\rho,\sigma))^2.
\eeq 
The above inequality helps us to overcome the drawbacks of relative entropy for not satisfying P4 and P5. Let us first rewrite  (\ref{sec1e}) in terms of trace distance as 
\beq\label{new2}
\bar{I}_{\Delta}^{T} \geq \max_{\rho_1(0),~\rho_2(0)} \left(D^T(\rho_1(0),\rho_2(0))-\left(\mathcal{N}_{\epsilon}^T(\bar{\rho}_1)+\mathcal{N}_{\epsilon}^T(\bar{\rho}_2)\right)\right).
\eeq
Using inequalities (\ref{new1}) and  (\ref{new2}), we arrive at 
\beq\label{sec1e1}
\bar{I}_{\Delta}^T \geq \max_{\rho_1(0),~\rho_2(0)} \left(D^T(\rho_1(0),\rho_2(0))-\left(\sqrt{\frac{\mathcal{N}_{\epsilon}^{Rel}(\bar{\rho}_1)}{2}}+\sqrt{\frac{\mathcal{N}_{\epsilon}^{Rel}(\bar{\rho}_2)}{2}}\right)\right),
\eeq
where $\mathcal{N}_{\epsilon}^{Rel}(\bar{\rho}_i)=S(\bar{\rho}_i||\rho_{th})$  denotes
the measure of non-ergodicity for the time-averaged state $\bar{\rho}_i$ in terms of relative entropy. 
As before, we can define a state-independent  measure of non-ergodicity as 
\beq\label{sec1e1a}
\mathcal{N}_{\epsilon}^{M(Rel)}=\max_{\rho(0)}S(\bar{\rho}||\rho_{th}), 
\eeq
The above defintion and the inequality (\ref{sec1e1}) leads to another uncertainty relation 
\beq\label{sec1e2}
\bar{I}_{\Delta}^T + \sqrt{2\mathcal{N}_{\epsilon}^{M(Rel)}}~ \geq \max_{\rho_1(0),~\rho_2(0)} \left(D^T(\rho_1(0),\rho_2(0))\right),
\eeq
in terms of trace and relative entropy distances. But it is to be noted that there is a certain limitation in this relation, because of the fact that the relative entropy is not a bounded function. When $\mbox{supp}~\rho \nsubseteq \mbox{supp}~\rho_{th}$, the relative entropy diverges. One such example is obtained for the zero temperature bath, where $\rho_{th}=|0\ket\bra 0|$ is pure. In that case, the relation (\ref{sec1e2}) becomes trivial. But in that case, we can find state- dependent uncertainty relations by defining state- dependent information loss as 
\beq\label{R2}
\bar{I}_{\Delta}(\bar{\rho}_1,\bar{\rho}_2) =  \left(D(\rho_1(0),\rho_2(0))-D(\bar{\rho}_1,\bar{\rho}_2)\right).
\eeq
This will lead us to the state-dependent uncertainty relation 
\beq\label{R3}
\bar{I}_{\Delta}^T(\bar{\rho}_1,\bar{\rho}_2)+\sum_{i=1,2}\sqrt{\frac{\mathcal{N}_{\epsilon}^{Rel}(\bar{\rho}_i)}{2}} \geq D^T(\rho_1(0),\rho_2(0)).
\eeq
But other than these extreme cases, the relation (\ref{sec1e2}) works perfectly.

Note that the distinguishability measures like trace distance, Bures distance and Jensen-Shanon divergence, mentioned earlier, not only satisfies all the conditions P1-P5, but they are also bounded. But in the cases of some unbounded distance measure, to avoid the triviality of the uncertainty relation (\ref{sec1eg}), we can use the state-dependent uncertainty relation 
\beq\label{R4}
\bar{I}_{\Delta}(\bar{\rho}_1,\bar{\rho}_2)+ \sum_{i=1,2}\mathcal{N}_{\epsilon}(\bar{\rho}_i)\geq D(\rho_1(0),\rho_2(0)).
\eeq

\subsection{Qubits} Upto now, we have considered an arbitrary density matrix of arbitrary dimension. Let us now restrict to the case of a two- level system (TLS) as a simple example to further understand the connection between non-ergodicity and information loss.
For a TLS, the pair of states maximizing the trace distance is located on the antipodes of the Bloch sphere i.e., the pair of  states consists of pure and mutually orthogonal states \citep{qubit}. 
Therefore in the case of trace distance, the uncertainty relation \eqref{sec1eg}, for a qubit, reads as
\beq\label{sec1fB}
\bar{I}_{\Delta}^{T} + 2\mathcal{N}_{\epsilon}^{M(T)} \geq 1.
\eeq
Similarly, the uncertainty relation given in \eqref{sec1e2} reduces to
\beq\label{sec1fa}
\bar{I}_{\Delta}^T + \sqrt{2\mathcal{N}_{\epsilon}^{M(Rel)}}~ \geq 1. 
\eeq
Let us now consider a simple Markovian model, where a qubit is weakly coupled with a thermal bosonic environment. In absence of any external driving Hamiltonian, the qubit eventually thermally equilibrates with the environment.
Under Born-Markov approximation, the master equation for this model is given by 
\beq\label{sec1fA}
\begin{array}{ll}
\dot{\rho}(\tilde{t})=\frac{i}{\hbar}[\rho(\tilde{t}),H_0]+\gamma(n+1)\left(\sigma_{-}\rho(t)\sigma_{+}-\frac{1}{2}\{\sigma_{+}\sigma_{-},\rho(\tilde{t})\}\right)\\
~~~~~~~+\gamma n\left(\sigma_{+}\rho(\tilde{t})\sigma_{-}-\frac{1}{2}\{\sigma_{-}\sigma_{+},\rho(\tilde{t})\}\right),
\end{array}
\eeq
where $H_0=\hbar\Omega_0|1\ket\bra 1|$ is the Hamiltonian of the system, $\gamma$ is a constant parameter and $n=1/(\exp(\hbar\Omega_0/K\tilde{T}_m)-1)$
is the Planck number. Here $\sigma_{+}$ and $\sigma_{-}$ are respectively the raising and lowering operators of the TLS,
with $|1\rangle$ being the excited state of the same. The solution of the Markovian master equation in \eqref{sec1fA} is given by 
\[\rho(\tilde{t})=\rho_{11}(\tilde{t})|1\ket\bra 1|+\rho_{22}(\tilde{t})|0\ket\bra 0|+\rho_{12}(\tilde{t})|1\ket\bra 0|+\rho_{21}(\tilde{t})|0\ket\bra 1|,\]
with
\[
\begin{array}{ll}
\rho_{11}(\tilde{t})=\rho_{11}(0)e^{-\gamma(2n+1)\tilde{t}}+\frac{n}{2n+1}\left(1-e^{-\gamma(2n+1)\tilde{t}}\right),\\
\rho_{22}(\tilde{t})=1-\rho_{11}(\tilde{t}),\\
\rho_{12}(\tilde{t})=\rho_{12}(0)\exp\left(-\gamma\frac{(2n+1)\tilde{t}}{2}-2i\Omega_0 \tilde{t}\right).
\end{array}
\]
One can find from the solution given above that the long time-averaged state for this evolution is independent of initial states and equal to the 
thermal state corresponding to the temperature of the bath $\tilde{T}_m$, which can be expressed as $p|0\ket\bra 0|+(1-p)|1\ket\bra 1|$, with $p=1/(1+\exp(-\hbar\Omega_0/K\tilde{T}_m))$. 
Hence the dynamics is ergodic and  we find that the information loss $\bar{I}_{\Delta}^T=1$; i.e. the system loses all its information. It is also noteworthy that Markovianity of a quantum evolution does not mean it will be ergodic. An example of such Markovian non-ergodic evolution is the dephasing channel expressed by the master equation 
\beq\label{R1}
\dot{\rho}= i\Omega_0[\sigma_z, \rho]+\gamma_d\left( \sigma_z\rho\sigma_z-\rho\right)
\eeq
Here the Lindblad operator is in the same basis as the system Hamiltonian $\sigma_z$. A system interacting with a bosonic environment can lead to such an evolution \citep{breuer}. The solution of this equation is given by 
\beq\label{R1A}
\begin{array}{ll}
\rho_{11}(\tilde{t})=\rho_{11}(0),\\
\rho_{22}(\tilde{t})=\rho_{22}(0),\\
\rho_{12}(\tilde{t})=\rho_{12}(0)e^{-2(i\Omega+\gamma_d)t}.
\end{array}
\eeq
We realize from Eq. (\ref{R1A}) that under this particular evolution, the system will decohere, but the digonal elements of the density matrix will remain invariant, leading to infinitely many fixed points for the dynamics. So this particular evolution will certainly be non-ergodic, since there exists infinitely many fixed points and the time averaged state will depend on the initial state of the system. This gives a definite example which proves that Markovianity does not imply ergodicity of the dynamics.
\section{Non-ergodicity and information back-flow in a central spin model}
In this section, we consider a specific non-Markovian model and study the status of uncertainty relation derived in Sec III. The system here consists of
 a single qubit, interacting with $N$ number of non-interacting spins. The total Hamiltonian of the system, governing the dynamics, is given by
\beq\label{sec2a}
\tilde{H}=\tilde{H}_S+\tilde{H}_B+\tilde{H}_I,
\eeq
where the system Hamiltonian $\tilde{H}_S$, bath Hamiltonian $\tilde{H}_B$, and interaction Hamiltonian $\tilde{H}_I$ are respectively given by 
\beq\label{sec2b}
\begin{array}{ll}
\tilde{H}_S=\hbar g\omega_0\sigma_z,\\
\tilde{H}_B=\hbar g\frac{\omega}{N}\sum_{i=1}^N\sigma_z^i,\\
\tilde{H}_I=\hbar g\frac{\alpha}{\sqrt{N}}\sum_{i=1}^N \left(\sigma_x\sigma_x^i+\sigma_y\sigma_y^i+\sigma_z\sigma_z^i\right).
\end{array}
\eeq
Here $\sigma_{k}$, $k=x,y,z$ are the Pauli spin matrices, the superscript `i' represents the ith spin of the bath, $g$ is a constant factor with the dimension of frequency, $\omega_{0}$ and $\omega$ are the dimensionless parameters characterizing the energy level differences
of the system and the bath respectively and $\alpha$ denotes the coupling constant of the system-bath interaction. By using the total angular momentum
operators $J_{k}=\sum\limits_{i=1}^N\sigma^{i}_{k}$, and the Holstein-Primakoff transformation, given by
\[  
J_+=\sqrt{N}b^{\dagger}\left(1-\frac{b^{\dagger}b}{2N}\right)^{1/2}~~,~~J_-=\sqrt{N}\left(1-\frac{b^{\dagger}b}{2N}\right)^{1/2}b,
\]
the bath and interaction Hamiltonians can now be rewritten as 
\beq\label{sec2c}
\begin{array}{ll}
\tilde{H}_B~~=-\hbar g\omega\left(1-\frac{b^{\dagger}b}{N}\right),\\
\tilde{H}_{I}=2\hbar g\alpha\left[\sigma_{+}\left(1-\frac{b^{\dagger}b}{2N}\right)^{1/2}b+\sigma_{-}b^{\dagger}\left(1-\frac{b^{\dagger}b}{2N}\right)^{1/2}\right]\\
~~~~~-\hbar g\alpha\sqrt{N}\sigma_z\left(1-\frac{b^{\dagger}b}{N}\right).
\end{array}
\eeq
We consider the initial (uncorrelated) system-bath state as $\rho_S(0)\otimes\rho_B(0)$.  Let us take the initial system qubit as $\rho_S(0)=\rho_{11}(0)|1\ket\bra 1|+\rho_{22}(0)|0\ket\bra 0|+\rho_{12}(0)|1\ket\bra 0|+\rho_{21}(0)|0\ket\bra 1|$ and the initial bath state to be a thermal state $\rho_B(0)=\exp(-\tilde{H}_B/K\tilde{T})$ in an arbitrary temperature $\tilde{T}$ with K being the Boltzmann constant. The reduced dynamics of the system state can then be calculated by tracing out the bath degrees of freedom and is given by 
$\rho_S(t)=\mbox{Tr}_B\left[\exp\left(-iHt\right)\rho_S(0)\otimes\rho_B(0)\exp\left(iHt\right)\right]$. Where
\[H=\frac{\tilde{H}}{\hbar g},~~t=g\tilde{t},~~\mbox{and}~~T=\frac{K\tilde{T}}{\hbar g},\]
are dimensionless, specifying Hamiltonian, time and temperature respectively. After solving the global Schr\"{o}dinger evolution, the reduced dynamics can be exactly obtained \citep{samya1,spin1} as 
\beq\label{sec2d}
\rho_S(t)=\left(\begin{matrix}
\rho_{11}(t) & \rho_{12}(t)\\
\rho_{21}(t) & \rho_{22}(t)
\end{matrix}\right),
\eeq
where
\beq\label{sec2e}
\begin{array}{ll}
\rho_{11}(t)=\rho_{11}(0)(1-\Theta_1(t))+\rho_{22}(0)\Theta_2(t),\\
\rho_{12}(t)=\rho_{12}(0)\Delta(t),
\end{array}
\eeq
with
\[
\begin{array}{ll}
\Theta_1(t)=\sum_{n=0}^N (n+1)\alpha^2(1-n/2N)\left(\frac{\sin(\eta t/2)}{\eta/2}\right)^2\frac{e^{-\frac{\omega}{T}(n/N-1)}}{Z},\\
\\
\Theta_2(t)=\sum_{n=0}^N n\alpha^2(1-(n-1)/2N)\left(\frac{\sin(\eta' t/2)}{\eta'/2}\right)^2\frac{e^{-\frac{\omega}{T}(n/N-1)}}{Z},\\
\\
\Delta(t)=\sum_{n=0}^N e^{-i(\Lambda -\Lambda')t/2}\left(\cos(\eta t/2)-i\frac{\theta}{\eta}\sin (\eta t/2)\right)\\
~~~~~~~~\times \left(\cos(\eta' t/2)+i\frac{\theta'}{\eta'}\sin (\eta' t/2)\right)\frac{e^{-\frac{\omega}{T}(n/N-1)}}{Z},\\
\\
Z=\sum_{n=0}^Ne^{-\frac{\omega}{T}(n/N-1)},\\
\\
\eta=2\sqrt{\left(\omega_0-\frac{\omega}{2N}-\alpha\sqrt{N}\left(1-\frac{2n+1}{2N}\right)\right)^2+4\alpha^2(n+1)(1-\frac{n}{2N})},\\
\\
\eta'=2\sqrt{\left(\omega_0-\frac{\omega}{2N}-\alpha\sqrt{N}\left(1-\frac{2n-1}{2N}\right)\right)^2+4\alpha^2 n(1-\frac{(n-1)}{2N})},\\
\\
\theta=2\left(\omega_0-\omega/2N+\alpha\sqrt{N}\left(1-\frac{2n+1}{2N}\right)\right),\\
\\
\theta'=-2\left(\omega_0-\omega/2N-\alpha\sqrt{N}\left(1-\frac{2n-1}{2N}\right)\right),\\
\Lambda=-2\omega\left(1-\frac{2n+1}{2N}\right)-\frac{\alpha}{\sqrt{N}},\\
\Lambda'=-2\omega\left(1-\frac{2n-1}{2N}\right)-\frac{\alpha}{\sqrt{N}}.
\end{array}
\]
The time-averaged state for this system can then be calculated as 
\beq\label{sec2f}
\begin{array}{ll}
\bar{\rho}_{11}=\rho_{11}(0)(1-\bar{\Theta}_1)+\rho_{22}(0)\bar{\Theta}_2,\\
\bar{\rho}_{12}=\rho_{12}(0)\bar{\Delta},
\end{array}
\eeq
with
\[
\begin{array}{ll}
\bar{\Theta}_1=\sum_{n=0}^N 2(n+1)\alpha^2(1-n/2N)\left(\frac{1}{\eta^2}\right)\frac{e^{-\frac{\omega}{T}(n/N-1)}}{Z},\\
\\
\bar{\Theta}_2=\sum_{n=0}^N 2n\alpha^2(1-(n-1)/2N)\left(\frac{1}{\eta'^2}\right)\frac{e^{-\frac{\omega}{T}(n/N-1)}}{Z},\\
\\
\bar{\Delta}=0.
\end{array}
\]
Note that in general the coherence of the time-averaged state will vanish as $\bar{\Delta}=0$. But there are specific resonance conditions under which there can be non-zero coherence present in the time-averaged state \citep{samya1}. But in this work, we will not consider such situations. 

Before investigating the uncertainty relation in terms of trace distance given in \eqref{sec1eg}, we explore the behavior of loss of information at instantaneous 
time with different parameters involved in this dynamics. For such study,
let us  restrict ourselves to the set of pure initial qubits over which the optimization involved in \eqref{sec1eg} is performed. In particular, we take the initial pair of orthogonal pure states to be $\cos\frac{\theta}{2}|1\ket + \sin\frac{\theta}{2}e^{-i\phi}|0\ket$ and $\sin\frac{\theta}{2}|1\ket - \cos\frac{\theta}{2}e^{-i\phi}|0\ket$, with $0\leq\theta\leq\pi,~ 0\leq\phi< 2\pi$. The instantaneous and average information losses in this case are given respectively by 
\beq\label{sec2g}
\begin{array}{ll}
I_{\Delta}^T(t)=\Theta_1(t)+\Theta_2(t), ~~\bar{I}_{\Delta}^{(T)}=\bar{\Theta}_1+\bar{\Theta}_2.
\end{array}
\eeq
\begin{figure}[htb]
	{\centerline{\includegraphics[width=7cm, height=6cm] {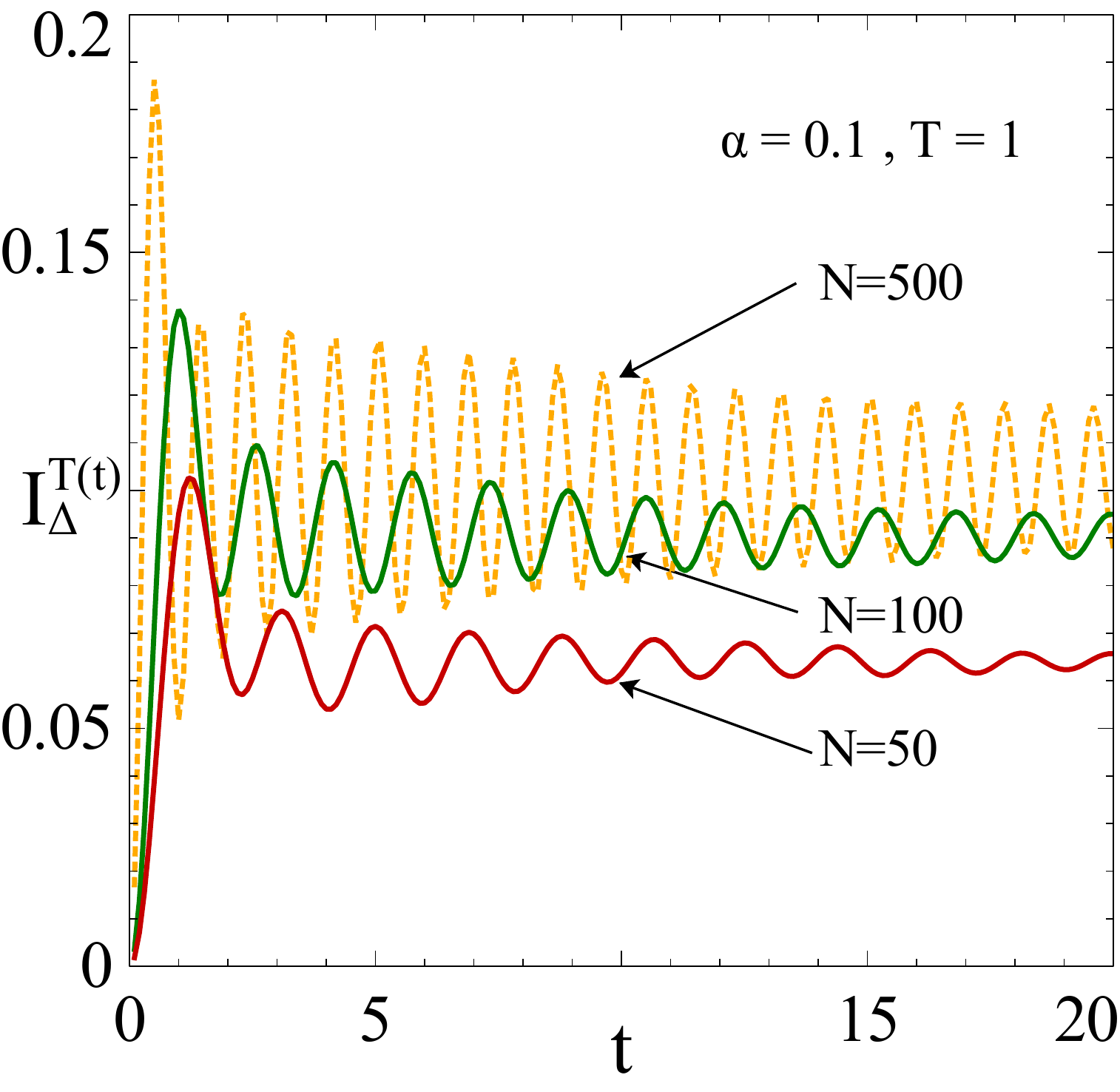}}}
	\caption{(Color online) Time-dynamics of instantaneous information loss. We plot $I_{\Delta}^T(t)$ on the vertical axis against t on the horizontal axis, for different values of the total number of bath spins $N$, where the system-environment duo governed by the Hamiltonian in Eq. \eqref{sec2a} is being considered. We set $\alpha=0.1$ and $T=1$. All quantities are dimensionless.  }
	\label{fig1}
\end{figure}
\begin{figure}[htb]
	{\centerline{\includegraphics[width=7cm, height=6cm] {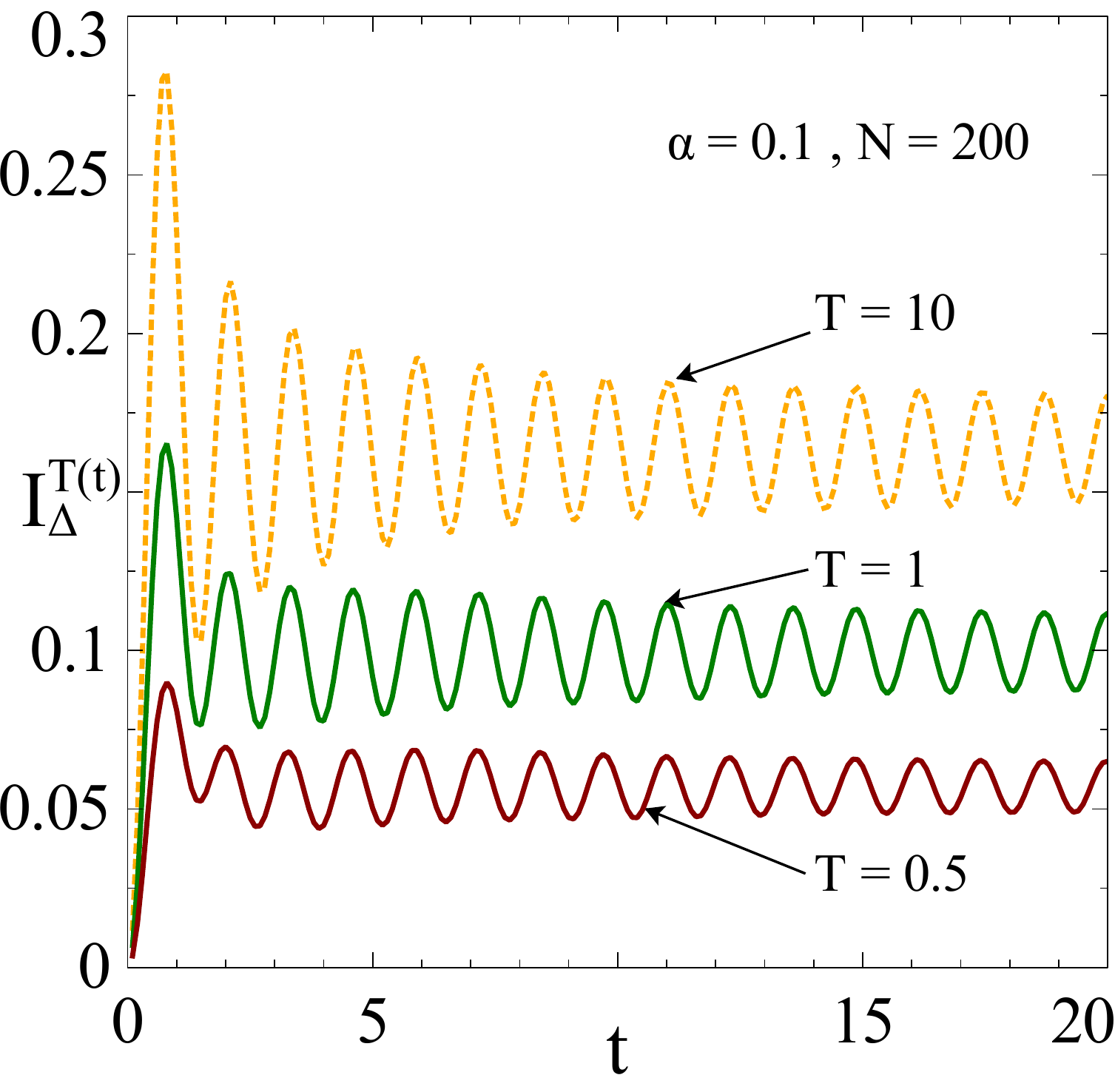}}}
	\caption{(Color online) $I_{\Delta}^T(t)$ vs $t$ for various temperatures. We set $N=200$ and $\alpha=0.1$. The physical system is the same as in Fig. \ref{fig1}. All quantities are dimensionless.}
	\label{fig2}
\end{figure}
\begin{figure}[htb]
	{\centerline{\includegraphics[width=7cm, height=6cm] {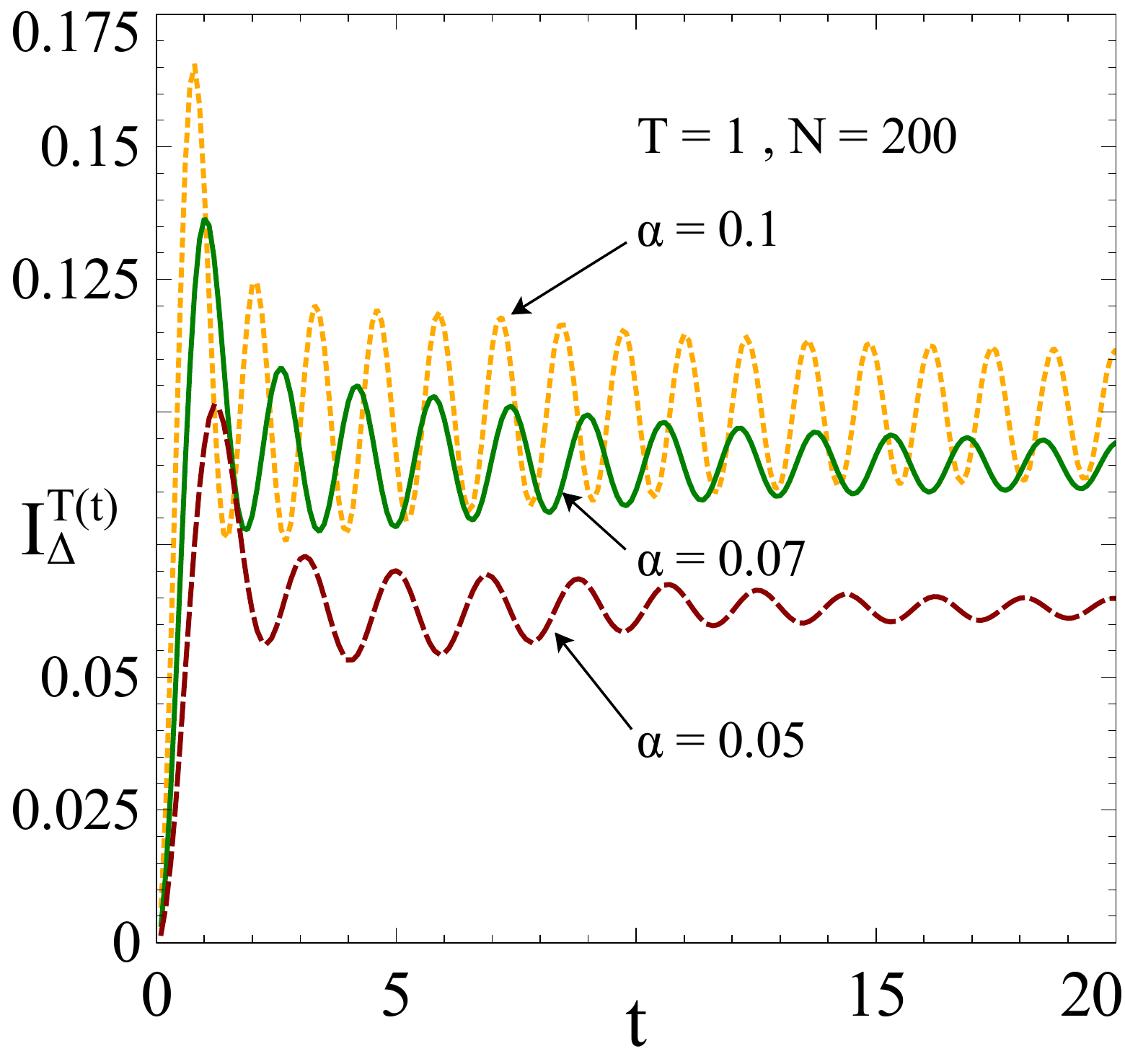}}}
	\caption{(Color online) $I_{\Delta}^T(t)$ with $t$ for three different values of system-bath coupling ($\alpha$). We set $T=1$ and $N=200$. The physical system is the same as in Fig. \ref{fig1}. All quantities are dimensionless.}
	\label{fig3}
\end{figure}
\noindent In Figs. \ref{fig1},\ref{fig2},\ref{fig3}, the instantaneous loss of information is depicted  with time for different values 
of the number of bath-spins ($N$), temperature ($\tilde{T}$) and system-bath interaction strength $(\alpha)$ respectively, by keeping other parameters fixed. From the figures,
we deduce the following: \\
\noindent\textbf{Observation 1:} The instantaneous loss of information shows oscillatory behavior whose amplitude decreases with time.\\
\noindent\textbf{Observation 2:} The increase of number of spins of the bath, in temperature, as well as in the interaction strength can be seen as increase 
of influence of bath on the system. Hence, expectantly in all cases, the loss of information increases  with increase of the above system parameters.

\begin{figure}[htb]
	{\centerline{\includegraphics[width=8cm, height=7cm] {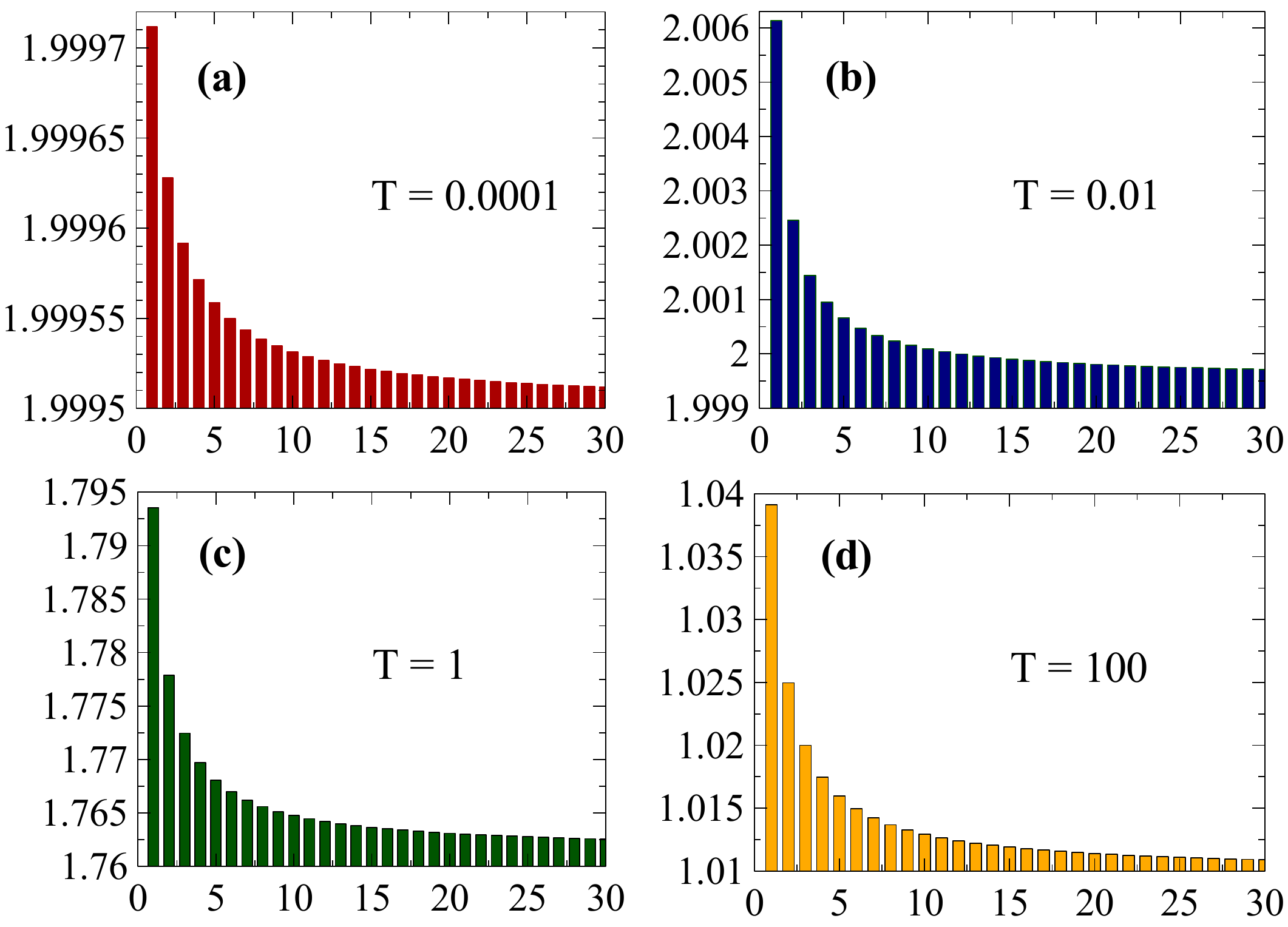}}}
	\caption{(Color online) Behavior of the uncertainty for different system-bath interaction strength. We denote the values of $I_{\Delta}^{T}+2\mathcal{N}_{\epsilon}^{M(T)}$ for different $\alpha$, as vertical bars on the horizontal axis that represents $\alpha$. Here, $N=1000$, and the different panels are for different $T$. The system considered is the one given by the Hamiltonian in Eq. \eqref{sec2a}.
	At already a moderate interaction strength, the quantity converges to a certain value, which is higher than unity. The converged value goes close to unity with the increase of temperature. All quantities are dimensionless.  }
	\label{fig5}
\end{figure}

Let us now check the uncertainty relation given in \eqref{sec1fB} for the qubit case, taking the same initial pair of pure orthogonal states and  the thermal state at arbitrary temperature $\rho_{th}=p_1|0\ket\bra 0|+(1-p_1)|1\ket\bra 1|$, where $p_1=\frac{1}{2}\left(1+\tanh\left(\frac{\hbar g\omega_0}{K\tilde{T}}\right)\right)$.
After performing the maximization, we find
\beq\label{sec2h}
\bar{I}_{\Delta}^{T}+2\mathcal{N}_{\epsilon}^{M(T)} =\bar{\Theta}_1+\bar{\Theta}_2+2|p_1-\bar{\Theta}_1|.
\eeq
We now examine the  conditions for which the uncertainty relation \eqref{sec1fB} saturates.
Note that for the ergodic situations, i.e. if the steady state is unique and is equal to the thermal state, the information loss is equals to unity, leading to a trivial  equality in \eqref{sec1fB}.
Keeping $N$ fixed to 1000 and fixing the temperature to different values,
we investigate the values of $\bar{I}_{\Delta}^{T}+2N_{\epsilon}^{M(T)}$ for increasing interacting strength. We observe that the sum goes close to
unity for a strong interaction strength as depicted in Fig. \ref{fig5} for high temperature. 

\begin{figure}[htb]
	{\centerline{\includegraphics[width=8cm, height=7cm] {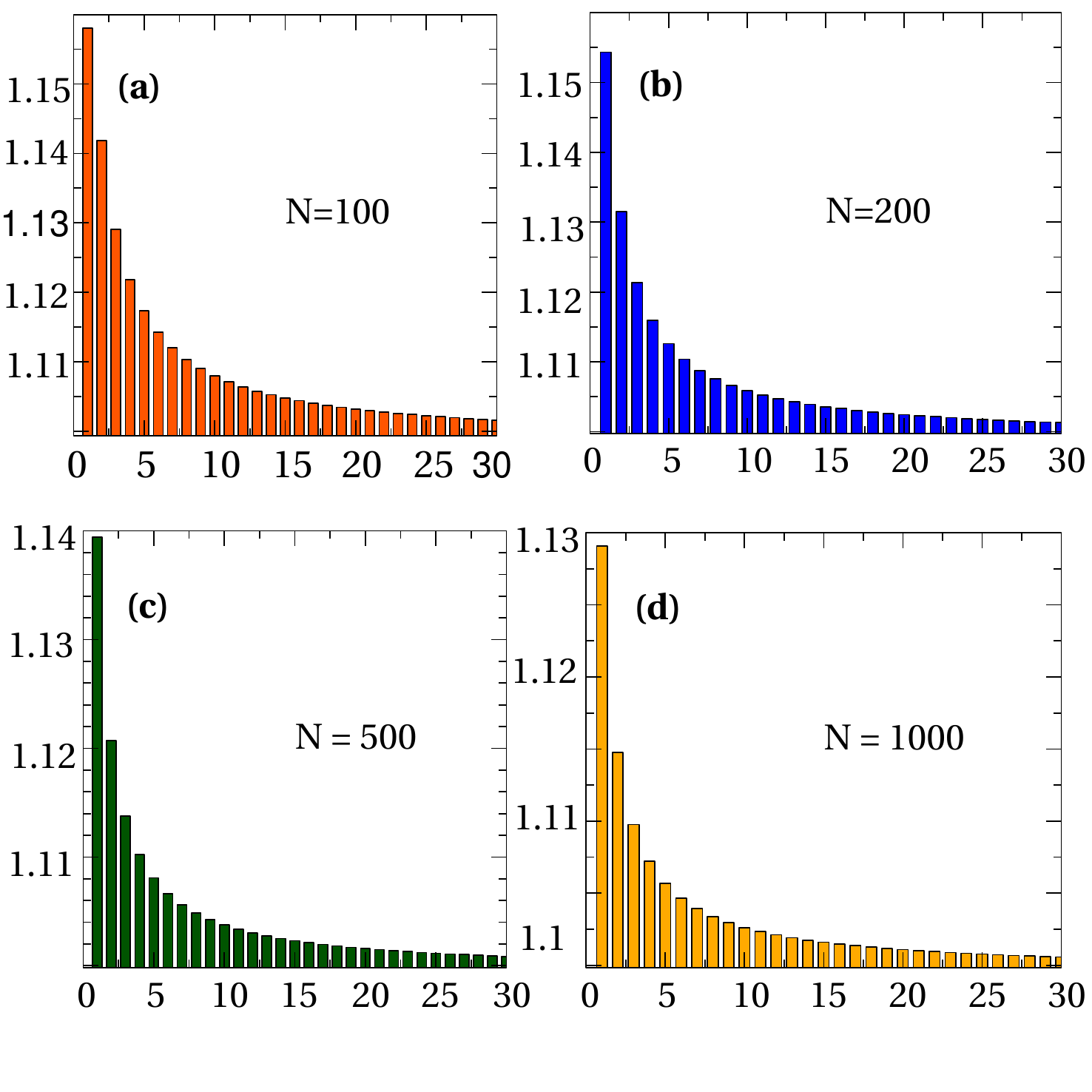}}}
	\caption{(Color online) Bar diagram for $I_{\Delta}^{T}+2\mathcal{N}_{\epsilon}^{M(T)}$
	against $\alpha$. The situation here is the same as in Fig. \ref{fig5}, except that the different panels are for different $N$, for fixed $T$ which is set to be 10. All quantities are dimensionless. }
	\label{fig6}
\end{figure}
\begin{figure}[htb]
	{\centerline{\includegraphics[width=8cm, height=7cm] {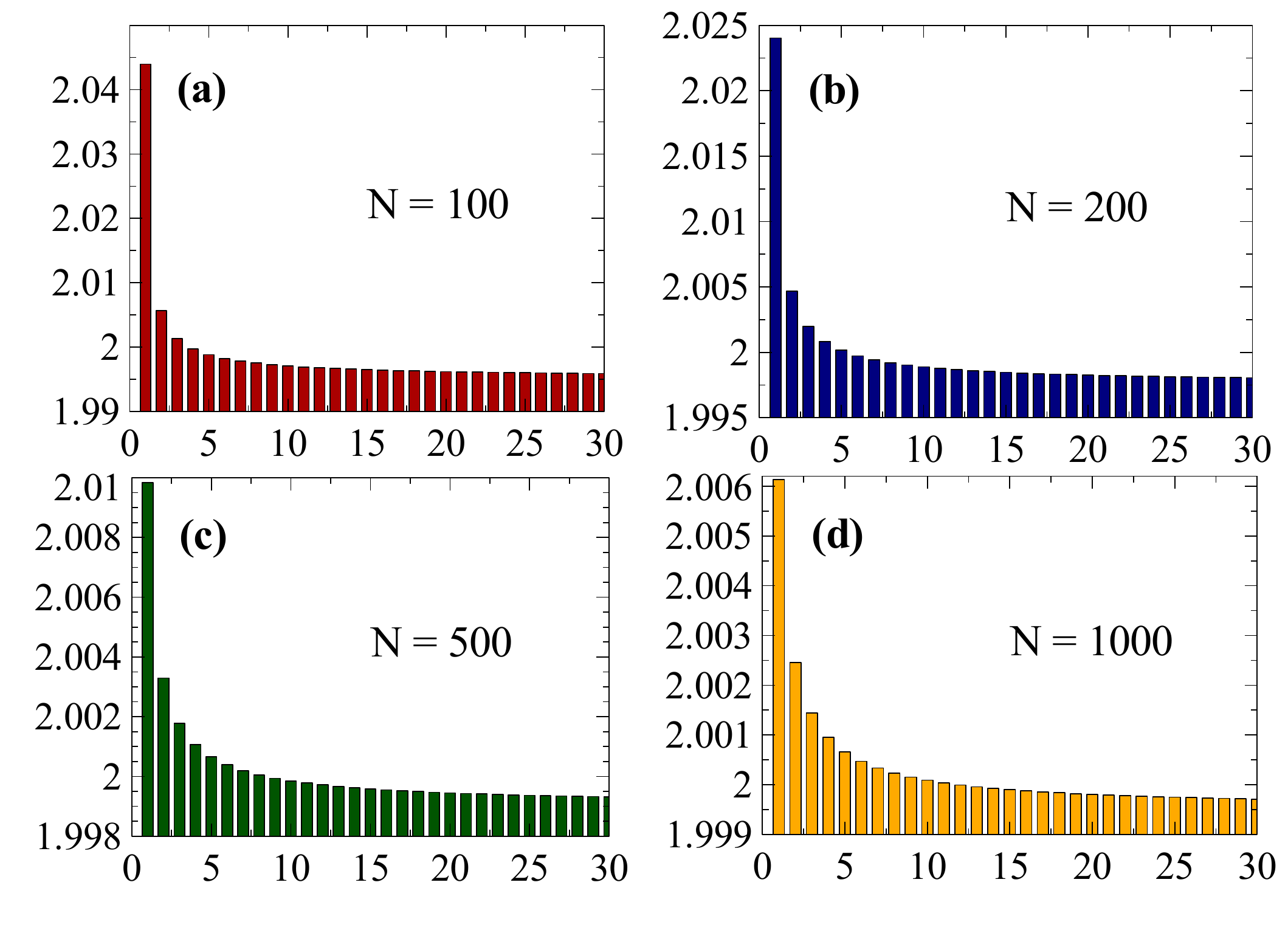}}}
	\caption{(Color online) The panels here are the same as in Fig. \ref{fig6}, except that $T=0.01$. All quantities are dimensionless. }
	\label{fig6a}
\end{figure}

\vspace{0.2cm}
\noindent In Figs. \ref{fig6} and \ref{fig6a}, we analyze the sum $I_{\Delta}^{T}+2\mathcal{N}_{\epsilon}^{M(T)}$ as the number of bath spins are ramped up from 100 to 1000. Scrutinizing these figures, we can safely conclude that with 
the increase in number of spins of the bath, or in the bath temperature, or in the system-bath interaction strength, the sum $I_{\Delta}^{(T)}+2\mathcal{N}_{\epsilon}^{M(T)}$  goes very close to unity in this qubit case, provided the optimization involved is restricted to pure qubits.
However, numerical evidence strongly suggests that for this non-Markovian model, given in Eq. \eqref{sec2a}, there will be no non-trivial situation when the 
uncertanity relation in Eq. \eqref{sec1fB} saturates to unity, provided the maximization is carried out over pure state.

We find that the saturation value of $\bar{\Theta}_1$ and $\bar{\Theta}_2$, when $\alpha\longrightarrow \infty$,  are respectively given by
\beq\label{sec2i}
\begin{array}{ll}
\bar{\Theta}_1^{sat}=\frac{1}{2}\sum_{n=0}^{n=N} \frac{1}{4+N
\frac{(1-(2n+1)/2N)}{(n+1)(1-n/2N)^2}}\frac{e^{-\frac{\hbar\omega}{KT}(n/N-1)}}{Z},\\
\bar{\Theta}_2^{sat}=\frac{1}{2}\sum_{n=0}^{n=N} \frac{1}{4+N
\frac{(1-(2n-1)/2N)^2}{n(1-(n-1)/2N)}}\frac{e^{-\frac{\hbar\omega}{KT}(n/N-1)}}{Z}.
\end{array}
\eeq  
If the interaction Hamiltonian given in Eq. \eqref{sec2b}, is considered in absence of the $z-z$ interaction, the saturated values of $\bar{\Theta}_1$ and $\bar{\Theta}_2$ in the limit $N\rightarrow\infty,~\alpha\rightarrow\infty$ will be $\bar{\Theta}_1^{sat}=\bar{\Theta}_2^{sat}=1/8$. In the infinite temperature limit, we have $p_1=1/2$, which leads to the equality in \eqref{sec1fB}. It is interesting that in the mentioned limit, the non-ergodicity measure is finite and equals to 3/8. Therefore, we find a non-ergodic situation, where the equality of the uncertainty relation holds.When the equality of the mentioned relations hold for a non-ergodic evolution, these relations imply that when the non-ergodicity of the dynamics increases, the information loss in the system decreases. Nonergodic dynamics are, in general, good for information processing as  they have less chance of leakage of  information compared to ergodic dynamics.
In particular, for non-ergodic evolution for which the uncertainty relations discussed in this paper are equalities, the loss of information can be quantified by and
attributed to the nonergodicity in the evolution.
It is also important to mention that spin bath models do not always indicate a shows non-ergodic dynamics. In a recent work \citep{markovSpin}, such dynamics hase been considered with the Born-Markov approximation region to find the effective reduced dynamics. It is shown in the mentioned work that there are situations where a unique fixed point (stationary state) can exist for the evolution, and hence in those situations, the dynamics is ergodic \citep{ergodicity3}. 

\section{Conclusion}
In open quantum dynamics, the information exchange between the system and bath plays an important role, while the time-evolved state's correspondence 
with the Gibb's ensemble conspire to imply the ergodic nature of the system. In this article, we establish a relation between loss of information and a measure of non-ergodicity. Both the definitions are given in terms of distinguishability, which can be measured by a suitably chosen distance measure. We have shown that the information 
loss and the quantifier of non-ergodicity follow  an uncertainty relation, valid for a broad class of distinguishability measures, which includes trace distance, Bures distance, Hilbert-Schmidt distance, Hellinger distance, and square root of Jensen-Shannon divergence.
We have further considered trace distance between a pair of quantum states as a specific distinguishability measure and connected the corresponding information loss with non-ergodicity, which is now defined  in terms of relative entropy between the time-averaged state and the thermal state, maximized over all possible initial states.
We have shown that in a Markovian model, the uncertainty relation saturates and shows a complete information loss. We also considered a structured environment model of a central quantum spin interacting,
according to Heisenberg interaction,
with a collection of mutually non-interacting quantum spin-half particles, leading to non-Markovian dynamics. In this case, we observed that with the increase of temperature, number of spins in the bath, and the system-bath interaction strength, there is increase in information loss
at instantaneous time. 
In this scenario, we found that the uncertainty
relation shows a nonmonotonic behavior with the increase of temperature for small values of interaction strength, provided the optimization is performed over pure qubits. Moreover, we found that although the uncertainty relation in this model goes close to the saturation value, it fails to saturate exactly. Interestingly  however, we found that in absence of $z-z$ system-bath interaction and in the limit of large bath size, high bath temperature, and strong system bath interaction, uncertainty relation between information loss and non-ergodicity, based on trace distance measure, is saturated, providing a non-ergodic situation that saturates the uncertainty. The uncertainty relations have been obtained by using the usual notion of the ergodicity where we require to have the unique fixed point, of the dynamics, to be thermal. We note that the entire analysis goes through for a more general definition, where a single fixed point is sufficient to imply ergodicity. 

\bibliographystyle{apsrev4-1}
\bibliography{ergodicity}

\end{document}